

\magnification=1200
\overfullrule=0pt
\baselineskip=20pt
\parskip=0pt
\def\dag{\dagger}
\def\del{\partial}

\def\a{\alpha}

\def\d{\delta}     \def\D{\Delta}
\def\e{\epsilon}

\def\q{\theta}     \def\Q{\Theta}

\def\m{\mu}	   
\def\n{\nu}        
\def\x{\xi}        
          
\def\p{\pi}        
\def\r{\rho}

\def\h{\chi}
\def\y{\psi}       

\def\yd{\y^{\dag}}

\def\br{\langle}
\def\ke{\rangle}
\def\ve{\vert}

\def\to{\rightarrow}

\def\zbar{\bar{z}}
\def\to{\rightarrow}

{\settabs 5 \columns
\+&&&&CCNY-HEP-93/2\cr
\+&&&&July 1993\cr}
\bigskip
\centerline{\bf $W_{\infty}$ GAUGE TRANSFORMATIONS AND THE ELECTROMAGNETIC }
\centerline{\bf INTERACTIONS OF ELECTRONS IN THE LOWEST LANDAU LEVEL}
\bigskip\bigskip
\centerline{ B. Sakita$^*$}
\bigskip

\centerline{ Physics Department, City College of the City University of New
York}
\centerline{ New York, NY 10031}
\bigskip
\centerline{\bf Abstract}
\bigskip
We construct a $W_{\infty}$ gauge field theory of electrons
in the lowest Landau level.
For this purpose we introduce an external gauge potential $\cal A $
such that its $W_{\infty}$ gauge transformations cancel against
the gauge transformation of the electron field. We then show that
the electromagnetic interactions of electrons in the lowest Landau
level are obtained through a non-linear realization of $\cal A$
in terms of the $U(1)$ gauge potential $A^{\m}$. As applications
we derive the effective Lagrangians for circular droplets
and for the $\n =1$ quantum Hall system.
\vfill

\noindent{$^*$E-mail address: sakita@sci.ccny.cuny.edu.}
\vfill\eject

The electromagnetic interactions of electrons in the lowest landau level
have been obtained recently from the original Lagrangian by
integrating out the components of electron field , which correspond to
the higher Landau levels [1].
The resulting effective Lagrangian does not have the
manifest gauge covariance that the original Lagrangian had. It appears that
the gauge covariance is maintained in the effective Lagrangian through a
non-linear
realization of $W_{\infty}$ gauge transformations\footnote*{This result could
have been
anticipated from the work of Shizuya [2], which deals with a general field
theoretic
formulation of electrons in a strong magnetic field.}.
In this paper we wish to elaborate
on this point and show that the effective Lagrangian can be {\it derived } to a
certain
extent as a non-linear realization of a $W_{\infty}$ gauge transformation.

Since the spin
of the electrons is aligned in the strong magnetic
field, we consider only a one component fermion field
$$
{\psi} (x,y,t) = {\sqrt{B\over {2\pi}}}\ e^{-{1\over 2}|z|^2 }\sum_{n=0}
^{\infty}
{{\bar{z} ^n} \over {{\sqrt {n!}}}}
{C}_n (t) \equiv  {\sqrt{B\over {2\pi}}}\
e^{-{1\over 2}|z|^2 }\sum_{n=0} ^{\infty}
\br z\ve n\ke
{C}_n (t)\ ,    \eqno (1)
$$
which obeys the lowest Landau level condition
$$
\left( \partial_z +{1\over 2}\bar{z}\right){\psi} (x,y,t)=0\ , \eqno (2)
$$
where $z = \sqrt{{B \over 2}}(x+iy)$,
$\bar{z} = \sqrt{{B \over 2}}(x-iy)$ and $\ve z\ke$ is the coherent state basis
of
a pair of bosonic operators $\hat{a}$ and ${\hat{a}}^{\dag}$.
The modes ${C}_n$ satisfy the usual anticommutation relations
$\{{C}_n, {C}^{\dag} _m
\}=\delta _{nm}$.

Next we consider a time dependent unitary transformation in the space
of ${C}_n$ [3]:
$$
{C}_n (t)\to {C'}_n (t)= u_{nm} {C}_m (t)=\br{n} \ve \hat{u}(t)\ve m \ke {C}_m
(t)\ . \eqno (3)
$$
An infinitesimal transformation
is generated by a hermititian operator which we write as
$\ddag \xi ( \hat {a} , {{\hat{a}}^{\dag}} ,t )\ddag$ with the anti-normal
order
symbol, where $\xi$ is a real function when $\hat{a}$ and ${\hat{a}}^{\dag}$
are replaced by $
z$ and $\bar z$ respectively. Then using (1) we obtain the following
infinitesimal transformation for ${\psi}$:
$$
\delta {\psi} (x,y,t) =-i{\sqrt{B\over {2\pi}}}\ e^{-{1\over 2}|z|^2 }
\ddag \xi ( \partial_{\bar{z}} , \bar{z} ,t )\ddag \sum_n \langle{z}\ve n\ke
{C}_n (t)
=-i\ddag \xi ( \partial_{\bar{z}}+{1\over 2}z , \bar{z} ,t )\ddag {\psi} (x,y
,t)\ .
\eqno (4)
$$
where $\ddag\ \ \ \ddag$ indicates that
the derivatives are
placed on the left of
$z$ and $\bar {z}$.
We call (4) the $W_{\infty}$ gauge transformation [3][4].

Introducing an external $W_{\infty}$ gauge potential $\cal A$
we write a $W_{\infty}$ gauge invariant
Lagrangian as follows\footnote*{To my knowledge this kind of
gauge transformation was first considered and used
by Das, Dhar, Mandal and Wadia in the context of $c=1$ matrix model [5].}:
$$
L=\int dx dy \bar{\psi}(x,y,t)\big(i\partial_t -V(x,y)-{\cal
A}(x,y,t)\big)\psi(x,y,t)\ ,
\eqno (5)
$$
where we have separated a confining static potential $V$ from $\cal A$.
The confining potential is a strong static potential provided by other
materials
outside the system,
and it confins the electrons to within the system.
Using (4) and (5) we obtain the following expression for the gauge
transformation of $\cal A$:
$$
\delta{\cal A} =\partial_t \xi +{1\over B}\{\!\!\{\xi , V\}\!\!\}
+{1\over B}\{\!\!\{\xi , {\cal A}\}\!\!\}\ , \eqno (6)
$$
where
$\{\!\!\{\ \  , \ \ \}\!\!\}$ is the Moyal bracket defined by
$$
\{\!\!\{\xi_1 ,\xi_2 \}\!\!\}=iB{\sum_{n=1} ^{\infty}}{{(-)^n}\over{n!}}\left(
{\partial_{z} ^{n}}\xi_1 {\partial_{\bar{z}} ^n}\xi_2 -
{\partial_{\bar{z}} ^{n}}\xi_1 {\partial_{z} ^n}\xi_2\right)\ , \eqno (7)
$$
which in the limit of large $B$ approaches to the Poisson bracket:
$$
{{\Longrightarrow}\atop{B\rightarrow\infty}}\ \ \{\xi_1 ,\xi_2\}_{P.B.}
=\epsilon_{0ij}\partial
^i\xi_1\partial^j\xi_2\ . \eqno (8)
$$
The gauge transformation (6) for large $B$ is then given by
$$
\delta{\cal A}(x,y,t) \approx n_{\mu}\partial^{\mu}\xi +
{1\over B}\epsilon_{0ij}\partial ^i\xi\partial^j {\cal A}\ , \eqno (9)
$$
where
$$
n_0 =1,\ \ \ \ \ n_i ={1\over B} \epsilon_{0ij}\partial ^j V \ . \eqno (10)
$$
The space vector $n_i$ is in general space dependent.

Let us imagine that a  weak space time dependent electromagnetic field
is further applied to the system. Let $A^{\mu}$ be its vector potential.
The Lagrangian of the system must be gauge invariant with respect to
$A^{\m}$, irrespective of the strong background electric and magnetic
fields. This implies that $\cal A$ and the $W_{\infty}$ gauge transformations
may be realized in terms of $A^{\m}$ and its gauge transformations.
Because $W_{\infty}$ gauge transformations (6) and (9) are of a non-abelian
type
,while the ordinary gauge transformations are abelian, non-linear realizations
are the only possibility.
We assume that the energy scale associated with $A^{\mu}$ is much smaller than
that of $V$ and $B$, so that we can neglect the higher derivatives of
$A^{\mu}$.
Thus it is sufficient to consider the transformation (9), which we should
realize by using
$$
\delta A^{\mu} = \partial^{\mu}\Lambda\ . \eqno (11)
$$

We notice first that since $\Lambda $ and $\xi$ are both infinitesimal $\xi$
should be linear in $\Lambda$. We expand $\cal A$ and $\xi$ as power series of
$A^{\mu}$ and its derivatives.
But by looking at the structure of equation (9) we
conclude that the following expressions are sufficient:
$$
\eqalign{{\cal A}=\, & a_{\mu}A^{\mu}+{1\over B}a_{\mu\nu}
A^{\mu}A^{\n}+{1\over B} a_{\mu\nu\sigma}
A^{\mu}\partial^{\nu}A^{\sigma}+ \cdots \cr
\xi=\, &\Lambda +{1\over B} \xi_{\mu\nu}\partial^{\mu}\Lambda
A^{\nu}+\cdots\cr}
\eqno (12)
$$
where $a$'s and $\xi_{\mu\nu}$ are space dependent parameters that are to
be determined.
We compute $\delta\cal A$ by using (11) and (12), and then use (9) to determine
$a$'s and $\xi_{\mu\nu}$. The results are
$$
\eqalign{
&\x_{[\m\n]}={1\over 2}(\x_{\m\n}-\x_{\m\n}),\ \ \
\x_{\{\m\n\}}={1\over 2}(\x_{\m\n}+\x_{\m\n}),
\ \ \ \xi_{[0i]}=0 ,\ \ \ \xi_{[ij]}=-{1\over 2}
\epsilon_{0ij}, \cr
&a_{\mu} = n_{\mu} + \kappa^{\alpha}\epsilon_{\alpha\mu \rho}
\partial^{\rho},\ \ \ a_{\m\n}
={1\over 2}\e_{0\m j}\del^j n_{\n} +{1\over 2}n_{\a}\del^{\a}\x_{\{\m\n\}},\cr
&a_{000}= \x_{\{00\}} ,\ \ \ a_{00i}=a_{i00}=\x_{\{0i\}} ,\ \ \
a_{0i0}=\x_{\{00\}} n_i ,\cr
&a_{ij0}=\epsilon_{0ij} +\x_{\{0i\}} n_j ,\ \ \ a_{i0j} =
-{1\over2}\epsilon_{0ij}
+{\xi}_{\{ij\}} ,\cr
&a_{0ij}=\x_{\{0j\}} n_i , \ \ \ a_{ijk}=\epsilon_{0ij}n_k
-{1\over2}\epsilon_{0ik}
 n_j +{\xi}_{\{ik\}} n_j ,\cr}\eqno (13)
$$
where constant $\kappa^{\alpha}$ and ${\xi}_{\{\m\n\}}(x,y)$ are
arbitrary. We remark that $n_i$'s are in general space dependent and
$\e_{0\m j}\del^j n_{\n}=\e_{0\n j}\del^j n_{\m}$ because of (10).
$\cal A$ is then given by
$$
\eqalign{{\cal A}=&\, n_{\mu}A^{\mu}
+{1\over B}\epsilon_{0ij}n_{\m}A^i \partial^j A^{\m} - {1\over
2B}\epsilon_{0ij}n_{\m}
A^i \partial^{\m} A^j+{1\over 2B}\epsilon_{0ik}({\del^k}n_{j})
A^i  A^j+
 \cr
+& \epsilon _{\alpha \mu\nu}\kappa^{\alpha}F^{\mu\nu}
+{1\over 2B}\epsilon_{0ij}
\partial^{\a}\big( n_{\a}\x_{\{\m\n\}}A^{\m}A^{\n}\big)\cdots
\cr}
\eqno (14)
$$
The first line of this expression does not contain the arbitrary constants.
It is the minimum realization up to the second power of $A^{\mu}$.
Let us denote it ${\cal A}_{\rm min}$. ${\cal A}_{\rm min}$ can be
further simplified:
$$
{\cal A}_{\rm min}=\, n_{\m}A^{\m}
+{1\over 2B}\epsilon_{\m \n \rho }A^{\m} \partial^{\n} A^{\rho}
 + {1\over 2B}\epsilon_{0ij}\partial^j\big(A^i n_{\m} A^{\m}\big) \ .\eqno (15)
$$
In the case of uniform background electric field,
$V=Ex$ and accordingly $n_i$ is a constant vector along the $y$ direction:
$$
n_i = v \delta_{iy},\ \ \ \  v\equiv{E\over B} \ \ \ {\rm for } \ \ V=Ex \
.\eqno (16)
$$
${\cal A}_{\rm min}$ is given by
$$
{\cal A}_{\rm min}=\, A^0 + vA^y
+{1\over 2B}\epsilon_{\m \n \rho }A^{\m} \partial^{\n} A^{\rho}
 + {1\over 2B}\epsilon_{0ij}\partial^0(A^i  A^j)
+{v\over 2B}\big( \partial^y (A^x A^y ) -\partial^x (A^y )^2\big) \ . \eqno
(17)
$$
This is precisely the expression obtained in [1].

As applications we consider briefly 1)
the edge fermions in a circular droplet ( a dot in a strong magnetic field)
and electromagnetic interactions
2) and the electromagnetic effective Lagrangian for a
$\n =1$ quantum Hall system.

\bigskip
\noindent{\bf Electromagnetic interaction of circular droplets
(dots in a strong magnetic field)}

Let $V(x,y)$ be a confining potential:
$$
V(x,y)={1\over 2}\a [(x^2 +y^2 ) - R^2 ]\ .\eqno (18)
$$
We compute the energy due to the confining potential. Using (1) we obtain
$$
\int dx dy \y^{\dag}(x,y)V(x,y) \y (x,y) = {\a\over B}\sum_{n=0}^{\infty}
 (n+1 -BR^2 /2)C^{\dag}_n C_n\ .
\eqno (19)
$$
The ground state is therefore obtained by filling the negative energy states:
$$
\ve G\ke =\prod_0 ^{N-1} C^{\dag}_n C_n\ve 0\ke\ , \eqno (20)
$$
where the total number of electrons $N$ is given by
$$
N={{BR^2}\over2}-{1\over 2}\ . \eqno (21)
$$
We keep only those electron operators that describe modes in the neighbourhood
of
the Fermi level. We define
$$
C_{N+n-1/2}=b_n ,\ \ \ v_\q =\a / B\eqno (22)
$$
With respect to the new vacuum (20), $b_n$ with positive $n$ is an annihilation
operator while with negative $n$, a hole creation operator.
(19) is then
$$
\int dx dy \y^{\dag}(x,y)V(x,y) \y (x,y)={\a\over B}\sum_n (n + {1\over 2} -N)
C^{\dag}_n C_n =\ \
v_\q\sum_{n={\rm half\ integer}}n \, b^{\dag}_n b_n\ .
\eqno (23)
$$
We restrict the range of sum to be $n\ll N$, although eventually
we will set $-\infty <n<\infty$ after taking the large $N$ limit.

We normal order the density operator with respect to the new vacuum. We obtain
$$
\r (x, y) \equiv \y^{\dag}(x,y) \y (x,y)=
{B\over {2\p}}\Q (r, R)+{B\over {2\p}}e^{-|z|^2} :\y^{\dag}(z)\y (\zbar ):\ .
\eqno (24)
$$
where
$$
\Q (r, R)=\ e^{-|z|^2}\sum _{n=0}^{N-1}{{|z|^2}\over n!}\ .\eqno (25)
$$
In the limit of large $B$ and $N$ with $R$ finite, we obtain
$$
\lim_{{B\to\infty}\atop{N\to \infty}}\Q(r, R) =\q(R-r)\ .\eqno (26)
$$
(see Appendix),
where $\q(r)$ is the standard step function.
Similarly we obtain
$$
\lim_{{B\to\infty}\atop{N\to \infty}}{B\over {2\p}}e^{-|z|^2}
:\y^{\dag}(z)\y (\zbar ):
={1\over R} \d (r-R) :\h^{\dag}(\q )\h(\q ): \ .\eqno (27)
$$
(see Appendix),
where
$$
\h (\q ) = {1\over\sqrt{2\p}}\sum _n e^{-in\q} b_n \ ,
\ \ \ \ \ z=|z|e^{i\q}\ .\eqno (28)
$$
Using (28) we can express
(23) as
$$
\int dx dy \y^{\dag}(x,y)V(x,y) \y (x,y)=\int d\q :\h^{\dag}(\q )
\big( -iv_\q \del_\q\big)\h (\q ):\eqno (29)
$$

Now it is straightforward to compute the effective Lagrangian (5). We obtain
$$
L= L_{\rm bulk}+L_{\rm boundary}\ ,\eqno (30)
$$
$$
L_{\rm bulk} =\int_{r,R} dx dy \big( - A^0 +{1\over{2\p}}{ b} +
{\cal L}_{C-S}\big)\ ,\eqno (30a)
$$
$$
L_{\rm boundary} =\oint d\q\Big[ \h^{\dag}\big((i\del_t -A_0 )-v_\q (i\del
_\q -A_\q )\big)\h -{{v_\q}\over {4\p}} A_\q ^2\Big]\ ,\eqno (30b)
$$
where
$$
{\cal L}_{C-S} =-{1\over{4\p}}\e_{0ij}\Big(A^i \del ^0
A^j - 2 A^i \del ^j A^0\Big) \eqno (31)
$$
and
$${ b}=\e_{0ij}\del^j A^i ,\ \ \ \ A_\q
=\e_{0ij}x^j A^i\ .\eqno (32)
$$

\noindent{\bf Electromagnetic effective Lagrangian for $\n =1 $ quantum Hall
system}

Let us consider a rectangular quantum Hall system of size
$L_x\times L_y \ , L_x\gg L_y$.
Neglecting the edges
of the $x$ ends, we can represent this system by choosing a confining
potential, which
depends only on $y$. The potential is flat in the middle and linearly rising at
the both ends.
Since in the limit of $A^\m =0$ the Hamiltonian is given by
${ H} =\int d^2 z e^{-|z|^2}\bar{\psi}(z)\psi(\bar{z})V(y)$
one can diagonilize it by choosing a $y$ diagonal basis for the LLL electron
field,
which we simply write as $\y (y)$ (the normalization is fixed by the
canonical anti-commutation relation): ${ H} =\int dy
\bar{\psi}(y)\psi(\bar{y})V(y)$.
This means that $V(y)$ is the dispersion of this effective one dimensional
system.
We adjust the chemical potential such that the zero energy Fermi level
intersects with $V$ precisely at $y=\pm {Ly\over 2}$.
The ground state is then the state
with all the negative energy single particle levels filled by the LLL
electrons.
The low energy excitations are then the motion of the electrons near
the Fermi level, i.e. electrons near the edges. These are edge excitations.
We make the electron density operator normal ordered with respect to the ground
state as before.
After the normal ordering we keep only these dgrees of freedom of
the right and left electrons near the Fermi level. The calculations are
entirely analogous
as before.
We obtain
$$
\rho (x, y)\approx {B\over 2\p}\left({\Theta}(y;-{{L_y}\over 2}, {{L_y}\over
2})
+\, e^{-|z_R |^2}:\bar{\y} _R (z_R )\y_R (\zbar_R ): +\, e^{-|z_L |^2}
:\bar{\y} _L (z_L )\y_R (\zbar_L ):
\right)\ , \eqno (33)
$$
where
$$\eqalign{&
{\Theta}(y;-{{L_y}\over 2}, {{L_y}\over 2})=
{\sqrt{B\over \p}}\int^{{L_y}\over 2}_{-{{L_y}\over 2}}
dy' e^{-B(y'-y)^2}\ {{\Longrightarrow}_{B\rightarrow\infty}}\ \
{\theta}(y-{{L_y}\over 2})\,{\theta}(y+ {{L_y}\over 2})\ ,\cr
&
\y_{R\atop L} (\bar{z}_{R\atop L})=\left({B\over \p}\right)^{1\over 4}
\int_{-{\D\over 2}} ^{\D\over 2}
dy' e^{-{1\over 2}(y'{}^2 B -|z_{R\atop L} |^2 )-
{\sqrt{2B}}iy'\bar{z}_{R\atop L}}
\y_{R\atop L} (y')\ ,\cr
&
z_{R\atop L}={\sqrt{B\over 2}}\big( x +i(y\mp {L_y\over 2})\big)\ ,\ \ \ \ \ \
\y_{R\atop L} (y)=\y (y \mp {L_y\over 2})\ .\cr}\eqno (34)
$$
Substituting these into the Lagrangian (5), where we keep only
${\cal A}_{\min}$,
we obtain the following effective Lagrangian after some calculations:
$$
L=L_{\rm bulk} + L_{\rm Rboundary}+L_{\rm Lboundary} \ ,\eqno (35)
$$
$$
L_{\rm bulk} = \int\int_D dx dy \Big( -A^0 + {1\over{2\p}}b +  {\cal
L}_{C-S}\Big)
\ , \eqno (35a)
$$
$$
L_{R\atop L}=\int dx \left( :\yd _{R\atop L} (x,t) \big( i \del _t -A^0
_{R\atop L}
\pm v_{R\atop L}(i\del _x + A^x _{R\atop L})\big)\y_{R\atop L} (x,t) : -
{1\over{4\p}}
v_{R\atop L} ( A_{R\atop L} ^x (x,t))^2 \right)\ . \eqno (35b).
$$
where we set $\D\to\infty$ and used $\y_{R\atop L} (x)$ for
the Fourier transform of $\y_{R\atop L} (y)$. When we use this Lagrangian,
we must assume $\y_{R\atop L} (x)$ is a slowly varyiing function.

The $x$ component of the current density can be derived from (35)
by taking a functional derivative with
respect to $A^x$. It consists of a bulk current and the edge currents.
The bulk current is the contribution from the Chern-Simon term and
only this contributes
to the Hall current in the present formulation [6].

Finally, we should remark that the boundary Lagrangians (30b) and (35b) are the
Lagrangians of
chiral fermions interacting with an external electromagnetic field.
The anomalies associated with these systems and their corresponding
bosonization
have been discussed by many authours, for example in [7]. Taking
these anomalies into account one proves that the Lagrangians (30) and (35) are
gauge invariant [8].

We also remark that calculations similar to these two
applications discussed above have been done recently by the CERN group [9].

\noindent{\bf Appendix}: Proof of (26) and (27)

Let $|z|^2 =x={B\over 2} r^2$
$$
\Q (r;R)=F(x,N)\equiv e^x\sum_{n=0} ^{N-1} x^{n} /n!\ ,\eqno (36)
$$
$$
{\del\over{\del x}}F(x,N) =-F(x,N)+F(x,N-1)=-e^{ -x}x^{N-1}/(N-1)!\ . \eqno
(37)
$$
On the other hand
$$
\y (\zbar )=\sum_{n=-N+1/2}^{\infty}{{(\zbar
)^{N+n-1/2}}\over{\sqrt{(N+n-1/2)!}}}b_n
\approx{{|z|^{N-1/2}}\over\sqrt{(N-1/2)!}}e^{-iN\q}\sqrt{2\p}\ \h (\q )\eqno
 (38)
$$
Thus,
$$
{B\over {2\p}}e^{-|z|^2} :\y^{\dag}(z)\y (\zbar ):\approx
Be^{-|z|^2}{{(|z|^2)^{(N-1/2)}}\over{(N-1/2)}}
:\h^{\dag}(\q )\h(\q ): \approx
-{\del\over{\del x}}F(x,N) B
:\h^{\dag}(\q )\h (\q ): \ .\eqno (39)
$$
We evaluate $e^{ -x}x^{N-1}$ by the saddle point method:
$$
e^{ -x}x^{N-1}=e^{-(x-(N-1){\rm ln}x)}\approx e^{-x_0 +
x_0{\rm ln}x_0} e^{-{1\over 2}
(x-x_0 )^2 /x_0 }\, \ \ \ \ \
x_0 = N-1\ .
$$
Using this and Sterling's formula
$
(N-1)!=x_0 ! \approx\sqrt{2\p} e^{-x_0}(x_0 )^{x_0 +1/2}
$
we obtain
$$
-{\del\over{\del x}}F(x,N)=-{1\over{\sqrt{2\p x_0}}}e^{-{1\over 2}
(x-x_0 )^2 /x_0 }\approx{1\over BR}{\sqrt{B\over\p}}e^{-(r-R)^2}
\Longrightarrow_{B\to\infty}{1\over{BR}}\d (r-R) \ , \eqno (40)
$$
where we used (21). From (39) and (40), (27) follows.
Notice $F(0,N)=1$ from the definition (36).
With this boundary condition we integrate (40) to obtain (26).

\centerline{\bf Acknowledgements}

We acknowledge useful discussions with,
R. Ray,  Z.-b. Su and especially with D. Karabali who
has also checked most of the calculations.
This work was supported by the NSF grant PHY90-20495 and the Professional
Staff
Congress Board of Higher Education of the City
University of New York under grant
no. 6-63351.

\bigskip
\centerline{\bf References}
\bigskip
\item{[1]}  R. Ray and B. Sakita "{\it The Electromagnetic Interactions
of Electrons in the Lowest Landau Level}" Ann. Phys. (N.Y.) (in press).
\item{[2]}  K. Shizuya, { Phys. Rev.} {\bf B45} (1992) 11143.
\item{[3]}  S. Iso, D. Karabali and B. Sakita, { Phys. Lett.}  {\bf B 296}
(1992) 143.
\item{[4]}  A. Capelli, C. Trugenberger and G. Zemba, { Nucl. Phys.}
{\bf B396} (1993) 465.
\item{[5]}  S. R. Das, A. Dhar, G. Mandal and S. R. Wadia; Int. J. Mod. Phys.
{\bf A7} (1992)
5165, Mod. Phys. Lett. {\bf A7} (1992) 937.
\item{[6]}  In this respect we cite a very recent preprint by Shizuya. K.
Shizuya "{\it
Gauge invariance and edge-current dynamics in the quantum Hall effect}"
YITP/U-93-8
\item{[7]}  J. Sonnenschein, { Nucl Phys.} {\bf B 309} (1988) 752.
\item{[8]}  X.G. Wen, { Phys. Rev.} {\bf B 43}, 11025, (1990)
\item{}  M. Stone, { Ann. Phys.} (N.Y.), {\bf 207} (1991) 38; { Int. J. Mod.
Phys.}
 {\bf B5} (1991)  509.
\item{}  J. Fr\"ohlich \& T. Kerler, { Nucl. Phys.} {\bf B354} (1991) 365.
\item{} See also [1].
\item{[9]} A. Cappelli, G. V. Dunne, C. A. Trugenberger and G. R. Zemba, "{\it
Conformal
Symmetry and Universal Properties of Quantum Hall States}" CERN-TH 6702/92.
\item{} A. Cappelli, C. A. Trugenberger and G. R. Zemba, "{\it Large N limit
in the Quantum Hall Effect}", CERN-TH 6810/93.

\end